\documentclass[%
 reprint,
superscriptaddress,
 amsmath,amssymb,
 aps,
prl,
]{revtex4-2}

\usepackage{graphicx}
\usepackage{dcolumn}
\usepackage{bm}
\usepackage{hyperref}

\begin{document}

\title{Electro-optic Fourier transform
chronometry of pulsed quantum light}

\author{Ali Golestani}
 \affiliation{Faculty of Physics, University of Warsaw, Pasteura 5, 02-093 Warszawa, Poland}
 
 \author{Alex O. C. Davis}
\email{aocd20@bath.ac.uk}
\affiliation{Centre for Photonics and Photonic Materials, Department of Physics, University of Bath, Bath BA2
7AY, United Kingdom}
\affiliation{Laboratoire Kastler Brossel, Sorbonne Université, ENS-Université PSL, CNRS, Collège de France, 4 Place Jussieu, F-75252 Paris, France}
 
\author{Filip So\'snicki}%
\author{Micha{\l} Miko{\l}ajczyk}
\affiliation{%
Faculty of Physics, University of Warsaw, Pasteura 5, 02-093 Warszawa, Poland}%

\author{Nicolas Treps}
\affiliation{Laboratoire Kastler Brossel, Sorbonne Université, ENS-Université PSL, CNRS, Collège de France, 4 Place Jussieu, F-75252 Paris, France}

\author{Micha{\l} Karpi\'nski}
\affiliation{Faculty of Physics, University of Warsaw, Pasteura 5, 02-093 Warszawa, Poland}

\date{\today}

\begin{abstract}
The power spectrum of an optical field can be acquired without a spectrally resolving detector by means of Fourier-transform spectrometry, based on measuring the temporal autocorrelation of the optical field. Analogously, we here perform temporal envelope measurements of ultrashort optical pulses without time resolved detection. We introduce the technique of Fourier transform chronometry, where the temporal envelope is acquired by measuring the frequency autocorrelation of the optical field in a linear interferometer. We apply our technique, which is the time-frequency conjugate measurement to Fourier-transform spectrometry, to experimentally measure the pulse envelope of classical and single photon light pulses.



\end{abstract}


\maketitle

Ultrashort optical pulses are essential for a variety of tasks \cite{weiner2011ultrafast}, from probing the dynamics of molecular systems \cite{Zewail2000} to precision metrology \cite{Hansch2002}. Recently there has been significant interest in using ultrashort pulses in low-light scenarios, most notably in quantum information science and technologies \cite{karpinski2021control} such as quantum computing \cite{Humphreys2013,Brecht2015,Lukens2017}, quantum communications \cite{Bob2013,Zhong2015,Leifgen2015,Islam2017} and quantum sensing \cite{Jian2012,Donohue2018}. 

For all applications, measuring the temporal intensity profile of a wave packet is a crucial characterisation capability. Temporal mode characterisation is especially critical for quantum information technologies, where losses induced by mode-mismatching can severely degrade the performance of quantum communication networks and optical information processors. The most well-developed technique, intensity autocorrelation, uses second harmonic generation to achieve direct sensitivity to the optical power \cite{diels2006ultrashort}. While nearly ubiquitous across classical ultrafast optics, this method suffers the serious drawback that the efficiency of the nonlinear generation vanishes with the optical intensity. Intensity autocorrelation is therefore inapplicable to low-intensity fields such as single-photon pulses, necessitating the development of alternative techniques.

Here, we devise and demonstrate a technique for measuring the temporal energy envelope of ultrashort single-photon pulses. Our approach, based on tunable electro-optic spectral shearing interferometery, is the time-frequency Fourier analogue of Fourier-transform spectrometry (FTS) -- one of the basic tools available in optical pulse characterisation \cite{bell2012introductory}. It can be viewed as interchanging the roles of time and frequency in a Fourier transfrom spectrometer. Furthermore, the method does not require optical nonlinearity, as in the case of autocorrelators. It employs electro-optic spectral shearing \cite{wright2017spectral,dorrer2003highly}, a deterministic linear optical process, and as such is applicable for fields with arbitrary intensity and photon number statistics. 

Previous work has delivered a variety of techniques to fully characterise low-intensity or single-photon fields, from which the intensity envelope can be recovered. However, these more comprehensive strategies, which also seek to determine the temporal phase, are not resource-efficient for the reduced problem of determining solely the temporal intensity. Our approach is fully linear optical and requires only non-resolving bucket detection of single photons. In this regard, the present work stands apart from spectral interferometry \cite{davis2018experimental,davis2018measuring,thiel2020single} which requires spectrally resolved detection of single photons, and externally referenced approaches that require a well-characterised classical field \cite{thekkadath2022measuring, Wasilewski2007} as well as nonlinear gating \cite{Ansari2016, maclean2018direct}. 

\begin{figure}[ht!]
\centering
\includegraphics[width=0.8\columnwidth]{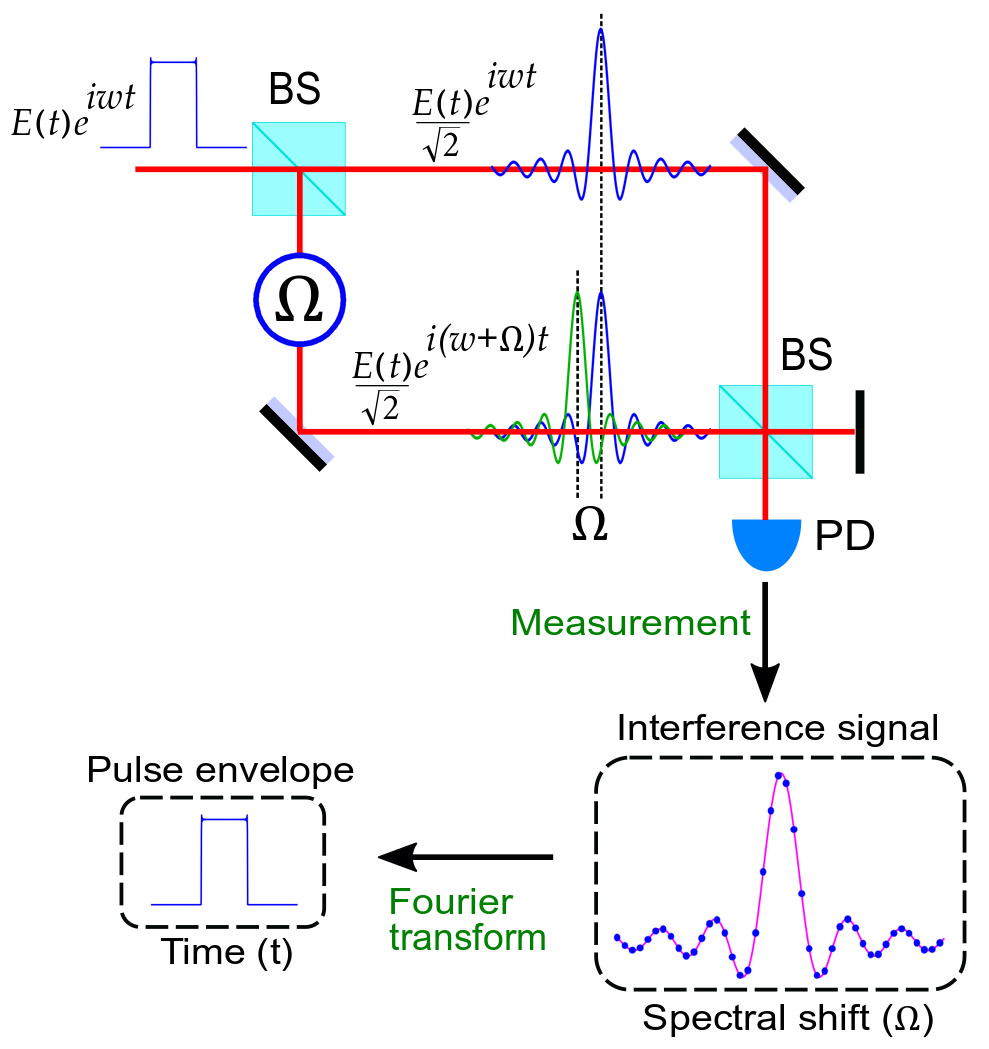}
\caption{Schematic of the Fourier transform tunable electro-optic spectral shearing interferometry. BS: Beam Splitter, PD: Photodiode\label{Figure1}}
\end{figure}

An overview of the experiment is shown in Fig. \ref{Figure1}. The optical pulse propagates through a balanced Mach-Zehnder interferometer where a tunable spectral shear $\Omega$ is applied to one arm of the interferometer. The central frequency of the optical pulse in this arm is scanned with respect to the field in the reference arm by applying an electro-optic spectral shear. At the output of the interferometer, the integrated energy per pulse is measured as a function of the spectral shear, whereby an interference pattern becomes apparent. The Fourier transform of the interference fringes directly yields the energy envelope of the optical pulse in the time domain.

The theoretical basis of our method closely parallels that of FTS, with time and frequency variables swapped and the role of the tunable delay substituted for tunable spectral shear in one arm of the interferometer.

In the co-moving picture, a pulsed optical mode, such as that occupied by a single photon, can be represented as a modulation of a carrier wave with frequency $\omega_0$ \cite{karpinski2021control}:
\begin{equation}
E(t)=\varepsilon(t)e^{i\omega_0(t-t_0)},
\end{equation}
where $\varepsilon(t)$ is a slowly-varying temporal mode function which we will assume is centred on $t=0$. A spectral shear can be imparted by the modulation of the field with an additional phase term that is linear in time, $e^{i\Omega (t-t_0)}$. In a refractive medium, this modulation can be achieved by a linear inclination in the index which co-propagates with the pulse \cite{wright2017spectral}.
The reference time $t=t_0$ is the zero crossing of this linear phase modulation: if this is displaced from the pulse centre of energy ($t_0\neq0$), the sheared pulse will also acquire a global phase offset of $\Omega t_0$ relative to the unsheared reference pulse. The effect of the spectral shear can then be represented as changing the frequency of the carrier wave, $\omega_0 \rightarrow \omega_0+\Omega$. 

The electric field at one output of the interferometer may therefore be written
\begin{equation}
E_\Omega(t)=\frac{1}{\sqrt{2}}\left(\varepsilon(t)e^{i\omega_0(t-t_0)}+\varepsilon(t)e^{i(\omega_0+\Omega)(t-t_0)}\right).
\end{equation}
The intensity at the output is then
\begin{equation}
I_\Omega(t)\equiv|E_\Omega(t)|^2=I_0(t)\left(1+\frac{1}{2}e^{i\Omega(t-t_0)}+\frac{1}{2}e^{-i\Omega(t-t_0)}\right),
\end{equation}
where $I_0(t)\equiv|\varepsilon(t)|^2$ has some characteristic duration $\Delta T$.
For single-picosecond and femtosecond pulses these timescales are too fast to be resolvable to single-photon detectors: the integration time for the detector is much slower than the envelope of the pulse. Hence all that is accessible is the integrated energy per pulse at the one output port of the interferometer as a function of $\Omega$:
\begin{align}
\mathcal{E}(\Omega)&\equiv \int I_\Omega(t) \mbox{d}t \\
&=\mathcal{E}_0+\frac{1}{2}\int \left[I_0(t)e^{i\Omega(t-t_0)}+I_0(t)e^{-i\Omega(t-t_0)}\right]\mbox{d}t \notag \\
&=\mathcal{E}_0+\frac{1}{2}e^{-i\Omega t_0}\mathcal{F}_t\{I_0(t)\}(\Omega)+\frac{1}{2}e^{i\Omega t_0}\mathcal{F}_t\{I_0(-t)\}(\Omega), \notag
\end{align}
where $\mathcal{E}_0\equiv\int I_0(t)\mbox{d}t$ and $\mathcal{F}_t\{\cdot\}(\Omega)$ and  $\mathcal{F}_\Omega\{\cdot\}(T)$ denotes the Fourier transform and its inverse respectively. 
We then calculate the inverse Fourier transform of $\mathcal{E}(\Omega)$ with respect to $\Omega$ to obtain:
\begin{align}
\bar{I}(T)&\equiv \mathcal{F}_\Omega\{\mathcal{E}(\Omega)\}(T) \label{seven}\\
&=\mathcal{E}_0 \delta(T)+\frac{I_0(T-t_0)}{2}+\frac{I_0(-T+t_0)}{2}\label{eight},
\end{align}
where $\delta(T)$ is the Dirac delta function. Hence if $t_0\gg\Delta T$, the final two terms in Eq. \ref{eight} form distinct side-peaks, so can be isolated and directly identified with the temporal intensity distribution $I_0(t)$ of the original pulse. The time reference $t_0$ determines the fringe spacing in the interferogram $\mathcal{E}(\Omega)$, and hence the peak separation in $\bar{I}(T)$.
Its analogue in FTS is the carrier frequency of the pulse.

Due to practical limits on the achievable spectral shear $\Omega$, the interferogram $\mathcal{E}(\Omega)$ can only be measured over some finite range $\Delta\Omega$, so the terms in Eq. \ref{eight} are convolved with a point-spread function with width $\sim 1/\Delta \Omega$. For the peaks to be distinct we therefore also require $t_0 \gg 1/\Delta\Omega$. For a maximum achievable spectral shear of hundreds of gigahertz, this condition requires $t_0$ of several picoseconds, i.e. a linear index modulation with its zero crossing some picoseconds away from the centre of the pulse. In practice it is difficult to sustain a single linear index ramp over such a duration. Fortunately, the temporal phase only needs to resemble a linear ramp (with some offset $\Omega t_0$) over the period where the pulse has significant intensity. We achieve this by locking the linear part of the ramp to the centre of the pulse and effecting the nonzero value of $t_0$ by applying a global phase offset proportional to $\Omega$ (modulo $2\pi$). 
Ideally, this would be a global offset phase that would be independent of frequency, e.g. by an achromatic phase shifter, but since the bandwidth is small relative to centre frequency (and the required phase shift restricted to the range {$0,2\pi$}) it is acceptable to simply use a delay line. This procedure can be thought of as controllably introducing the fringes necessary to isolate the interference terms in Eq. \ref{eight} from one another.

An optical pulse from a femtosecond laser (Menlo  Systems  C-Fiber  HP,  repetition  rate  of 80  MHz) at 1560~nm is taken and directed into a 4-\textit{f} spectral filter to prepare a pulse with 0.5~nm bandwidth. The pulse is prepared with diagonal polarisation and directed into a polarisation Mach-Zehnder interferometer \cite{davis2018experimental}. In such a scheme, the horizontal and vertical polarisation modes correspond to the principal axes of a birefringent optical path including an electro-optic phase modulator (EOPM) and polarisation maintaining fibres, thus constituting the two spatially overlapping arms of a Mach-Zehnder interferometer. The two arms accumulate a relative delay which is compensated by an appropriate length of birefringent material in the form of a length of PM fibre and a Soleil-Babinet compensator. A beam pickoff directs some of the optical signal from the laser output to a fast photodiode (EOT ET-3500F), the output of which is amplified and used to provide the radio-frequency (RF) driving signal for the EOPM \cite{poberezhskiy2003electro,sosnicki2020aperiodic}. A variable delay line, half-wave plate (HWP) and a polarising beam splitter (PBS) are used before the fast photodiode to control the optical intensity incident on it. This in turn controls the amplitude of the RF driving signal for the EOPM and hence enables the tunability of the spectral shear (\ref{Figure1}).

\begin{figure}[ht!]
\centering
\includegraphics[width=1\columnwidth]{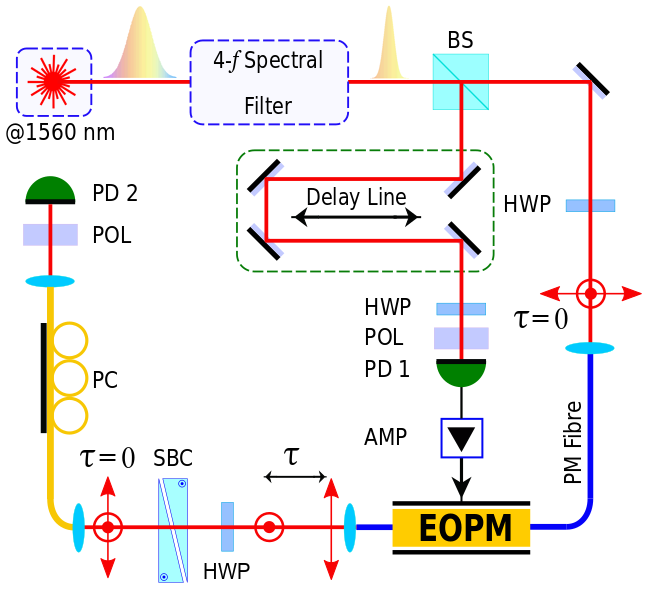}
\caption{Optical setup. BS: Beam Splitter, HWP: Half-Wave plate, POL: Polariser, PD: Photo-diode, EOPM: Electro-Optic Phase Modulator, AMP: Amplifier, SBC: Soleil-Babinet Compensator, PC: Polarisation Controller, $\tau$: Delay between polarisation components \label{Figure2}}
\end{figure}

The optical pulse must be shorter than the time duration of the RF electronic signal driving the EOPM, since it is necessary to have a uniform linear temporal phase ramp over the full pulse duration. The fast photodiode and amplifier provide a signal with a linear ramp of $\sim25$~ps duration. This defines the maximum temporal width of the optical pulse that can be sent into the EOPM without significant distortion due to higher-order temporal phase modulation. The change of slope and hence $\Omega$ changes the spectral shift but does not define an appropriate value of $t_0$. A suitable $t_0$ is emulated by applying an additional phase shift $\delta\phi \equiv t_0\delta\Omega$ with each increment of the shear $\delta\Omega$. This is done using a phase shifter incorporated into the interferometer, the Soleil-Babinet compensator. This procedure emulates the temporal phase modulation $e^{i\Omega(t-t_0)}$ over the duration of the pulse, and ensures that the interference pattern $\mathcal{E}(\Omega)$ features sufficiently dense interference fringes to distinguish the interference terms in Eq. \ref{eight} and hence recover $I_0(t)$.

The interferometer is initialised in its balanced configuration and the optical pulse is placed in the centre of the linear part of the electronic RF signal using the fibre-coupled variable delay line.  After each spectral shift increment, a global constant phase shift which is proportional to the spectral shift is applied using the phase shifter. The integrated energy per pulse at one output port of the interferometer is measured using a photodiode and recorded by an oscilloscope (Tektronix DPO7254C). This process is sequentially repeated until the spectrally sheared pulse is clear of its replica so that there is almost no overlap between the two arms of the interferometer in the spectral domain.

\begin{figure*}[ht!]
\centering
\includegraphics[width=2\columnwidth]{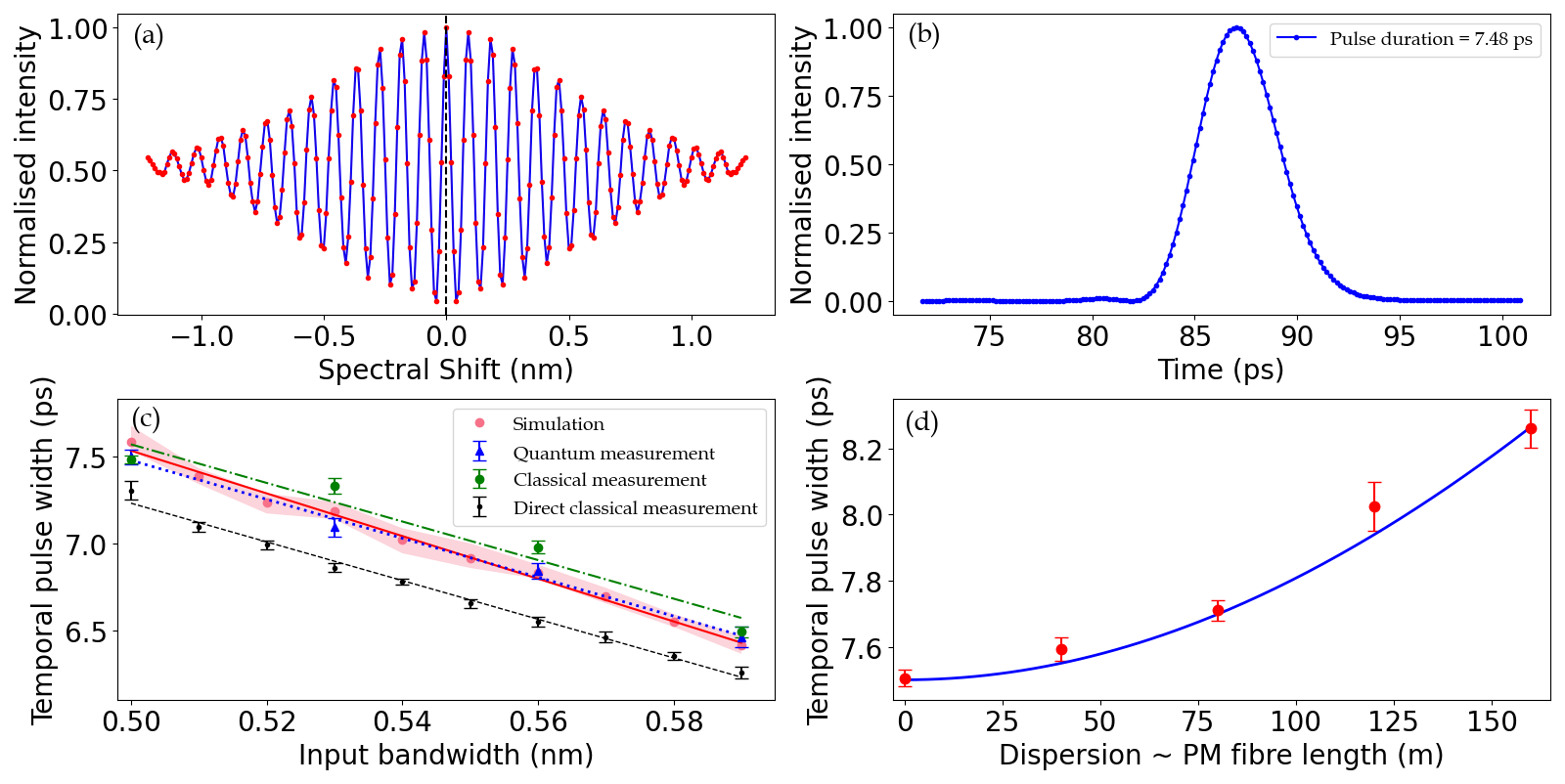}
\caption{(a) Measured interference signal as a function of spectral shift (red points) in which the positive side has been measured and the negative side is the mirror of the measured data. The blue line is used to guide the eye. (b) Energy envelope of the retrieved optical pulse in temporal domain. (c) Impact of input bandwidth change on temporal width of the classical (green points) and quantum (blue points) light pulse -- experimental data along with linear fits as shown by green dash-dotted and blue dotted lines respectively. Red points: theoretical simulation.
The temporal pulse width is calculated based on second-moment method. The red shaded area shows the variation of temporal pulse width for up to 10 percent error in the applied spectral shift. The direct measurement of the temporal duration of the classical pulses using the Optical Complex Spectrum Analyzer is shown by black points along with linear fit as indicated by the black dashed line. (d) Measured temporal width (red points) of the optical pulse versus dispersion (added PM fibre). The blue line is the theoretical calculation for the measured initial temporal width of the optical pulse before adding dispersion. \label{Figure3}}
\end{figure*}

We first demonstrate our technique with classical laser pulses with input spectral bandwidth of $0.50$~nm. A total spectral shear of $\Delta\Omega=1.22$~nm was applied to the optical pulse in one arm of the interferometer, in sequential steps of $0.01$~nm, to scan the optical pulse with respect to its replica in the other arm of the interferometer. Figure \ref{Figure3}(a) shows the measured intensity at the output of the interferometer as function of spectral shear. The fringe visibility decreases with increasing spectral shift due to the reduction in overlap in the spectral domain. The spectral shift was performed only in one direction (blue shift)  resulting in a fringe pattern for half of the optical pulse. We mirror this pattern to obtain the full symmetric fringe pattern as illustrated in Figure \ref{Figure3}(a). The temporal energy envelope of the optical pulse is achieved by taking a Fourier transform of the interference fringes with results shown in the Figure \ref{Figure3}(b). The calculated temporal width of the optical pulse is $7.48\pm0.02$~ps which is relatively close to theoretical expectations for the optical pulse with $0.5$~nm spectral bandwidth calculated based on second moment calculation.  

The measurement was performed over a range of  input bandwidths, measuring the change in temporal duration of classical optical pulses as indicated by the green points in Figure \ref{Figure3}(c). Due to the limitation in spectral shift (about 1.2 nm), it is not possible to take a larger input spectral width with the present setup. The results indicate that increasing the input bandwidth leads to the expected decrease in temporal duration. The measurement was repeated four times to consider the error in temporal width measurement. The theoretical simulation of pulse duration measurement was done using experimental input spectra and experimentally determined (by direct electro-optic sampling technique \cite{sosnicki2020aperiodic}) temporal phase profile that is depicted by red points in the Figure \ref{Figure3}(c). We also directly measured the temporal pulse duration of the classical pulses using Optical Complex Spectrum Analyzer (APEX Technologies AP2681A) as shown by black points in Figure \ref{Figure3}(c). The difference in the measured temporal width is to a large extent due to the residual distortion of the optical pulse spectrum when the spectral shear is applied, as confirmed by the theoretical simulation.  

To verify the validity of our measurement, the impact of adding dispersion to the pulse is investigated. Pulses of identical spectral bandwidth are sent through a variable length of optical fibre which leads to pulse broadening in the temporal domain without change in the spectral domain. The increase in temporal width corresponds to the amount of added dispersion. This measurement was done for added lengths of PM fibres in the optical setup. We considered the PM fibre length of $40$~m, $80$~m, $120$~m and $160$~m and each case was repeated four times in order to determine uncertainty in measuring the temporal width of the optical pulse.

As shown in Figure \ref{Figure3}(d), the result indicates that adding dispersion leads to increase in temporal width of the optical pulse. This increase follows a nonlinear trend for small amount of dispersion (less than about 1 km of SMF). Considering the formula, $T(z) = T_0 \sqrt{1+(\frac{z}{L_D})^2}$, we can calculate the temporal width at given point, $z$, after propagating along optical fibre \cite{agrawal2000nonlinear}. In this formula, $L_D = \frac{T_0^2}{|\beta_2|}$ where $\beta_2$ is group velocity dispersion and $T_0$ is the initial temporal duration of the optical pulse. In this calculation, the initial temporal width of the optical pulse was taken equal to 7.48 ps (the measured temporal width without added dispersion). The measured temporal widths after adding PM fibre (dispersion) in the setup are consistent with the theoretical model. Note that no fitting has been done in the theoretical calculation. 

\begin{figure}[ht!]
\centering
\includegraphics[width=1\columnwidth]{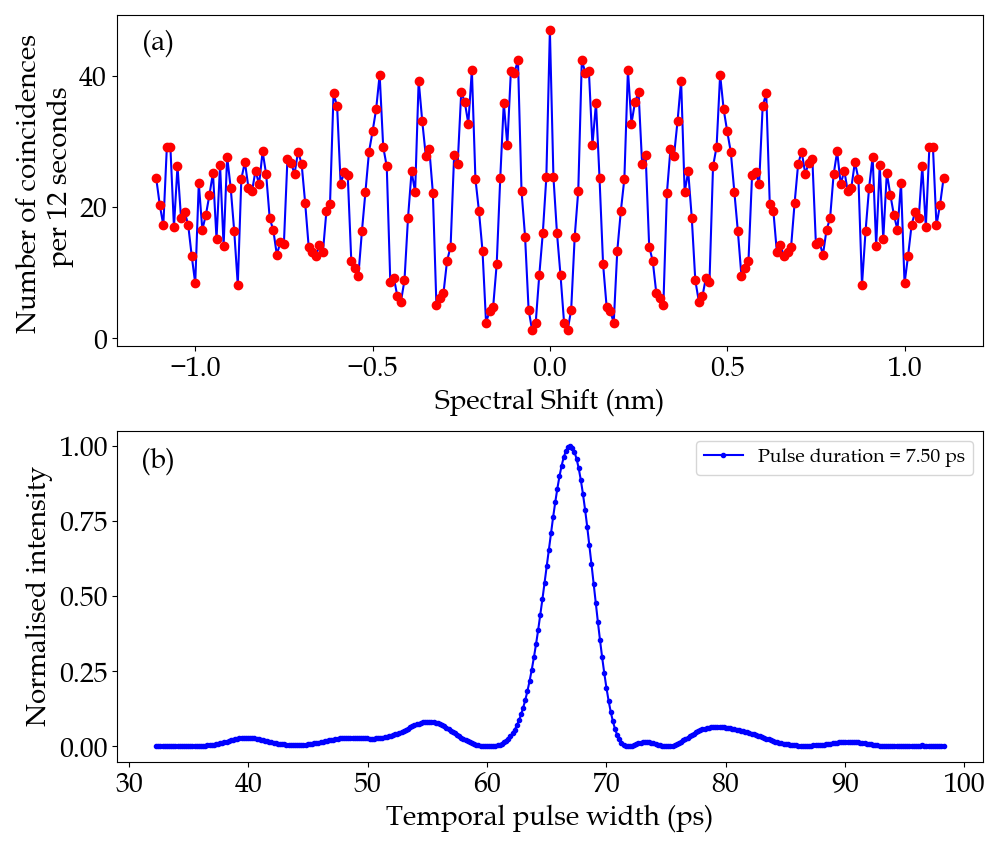}
\caption{a) Fringe pattern resulting from single-photon coincidences versus spectral shift where negative side is the mirror of positive side). The blue line is a guide to the eye. b) Energy envelope of the measured single photon pulse in temporal domain.\label{Figure4}}
\end{figure}

We then demonstrate the capability of the method to measure the temporal duration of heralded single photon pulses. To generate heralded photons, the optical pulses (central wavelength of $1560$~nm) from an erbium femtosecond oscillator (Menlo Systems C-Fiber HP 780, repetition rate of $80$~MHz) are frequency-doubled by a built-in second harmonic generation module. The up-converted optical pulses are spectrally filtered and focused into a bulk $10$-mm-long periodically poled potassium titanyl phosphate (PPKTP) nonlinear crystal where orthogonally polarised frequency degenerate photon pairs are generated through a type-II collinear spontaneous parametric downconversion process \cite{evans2010bright}. The orthogonally polarised photon pairs are coupled into a single mode polarisation maintaining fibre and split using a fibre polarisation beam splitter. One of the photons is sent into the fibre-pigtailed home-build 4-\textit{f} line where its bandwidth is reduced to 0.5 nm and then goes through the polarisation-based Mach-Zehnder interferometer. The other photon is directed straight to a single photon detector to act as a herald. Both photons are detected with niobium nitride superconducting nanowire single photon detectors (SNSPDs, Single Quantum) and time-tagged for time-resolved coincidence counting using a time-to-digital converter (Swabian Instruments Time Tagger Ultra).
Figure \ref{Figure4}(a) shows the measured coincidences as a function of spectral shear. Here, experimental uncertainties were estimated by bootstrapping assuming Poissonian photo-count statistics.

For this experiment, the number of coincidences was measured over a coincidence window of $10$~ns to count photons in a period less than the repetition rate ($12.5$~ns) of the pulsed laser. Preparing the input bandwidth of $0.5$~nm required a significant amount of spectral filtering that significantly decreased the number of coincidences, so an acquisition time of 12 seconds per data point was used for photon counting. The Fourier transform of the resulting interference fringes was taken and the temporal envelope of the single photon is shown in Figure \ref{Figure4}(b).

From the temporal profile, we calculate a temporal width of $7.50\pm0.04$~ps for the single photon which is close to the measured temporal width of the classical optical pulse. The measurement was done for a range of input bandwidth to see the temporal width change of the single photon pulse. The blue points in Figure \ref{Figure3}(c) indicate the expected decrease in temporal duration of the single photon pulse while the input spectral bandwidth is increased. This trend matches with results of the classical measurement as shown in Figure \ref{Figure3}(c).

In summary, the energy envelope of ultrashort classical and quantum light pulses was measured using Fourier-transform chronometry. It is a linear optical technique based on tunable electro-optic spectral shearing interferometry where the pulse temporal envelope is determined by measuring frequency dependent autocorrelation. In analogy to FTS, our technique can be viewed as a new photonic application of the well-established Wiener-Khinchin theorem \cite{wiener1930generalized, khintchine1934korrelationstheorie}. The measured value for the temporal width of the optical pulse is within $3$ percent of independently measured temporal duration, with the difference due to spectral broadening associated with spectral shear. The technique was validated by investigating the effect of bandwidth modification and added dispersion on the measured pulse duration. In both cases a very good agreement with theoretical predictions was obtained. Our method is central wavelength independent and intrinsically free of photon noise, making it an attractive tool for low light level applications. It may also enable classical pulse duration measurement at exotic central wavelengths. We expect that the range of pulse durations that can be measured will be greatly increased by means of thin-film lithium niobate technology for electro-optic phase modulation \cite{zhang2021integrated, zhu2021spectral}, possibly combined with complex temporal phase modulation \cite{johnson1988serrodyne, sosnicki2018large}. 

\

The experimental part of this work was carried out within the First TEAM programme of the Foundation for Polish Science (project no.\ POIR.04.04.00-00-5E00/18), co-financed by the European Union under the European Regional Development Fund. A part of this work was supported by the National Science Centre of Poland (project no. 2019/32/Z/ST2/00018, QuantERA project QuICHE).

\bibliographystyle{apsrev4-2}

\end{document}